%% file: main.tex
\documentclass[preprint,12pt,authoryear]{elsarticle}

\let\today\relax
\makeatletter
\def\ps@pprintTitle{%
    \let\@oddhead\@empty
    \let\@evenhead\@empty
    \def\@oddfoot{\footnotesize\itshape
         {Working Paper.} \hfill\today}%
    \let\@evenfoot\@oddfoot
    }
\makeatother

\usepackage[utf8]{inputenc}
\usepackage{verbatim}
\usepackage{amsfonts}
\usepackage{amssymb,amsmath,amsthm}
\usepackage{array, booktabs, makecell}
\usepackage{xcolor}
\usepackage{graphicx}
\usepackage{url}
\usepackage{multicol}
\usepackage{multirow}
\usepackage{subcaption}
\usepackage{natbib}
\usepackage{pifont}
\usepackage{subfiles}
\usepackage{soul}
\usepackage{threeparttable}
\usepackage[english]{babel}
\usepackage[nolist]{acronym}
\usepackage{hyperref}

\newtheorem{hyp}{Hypothesis} 
\newtheorem{subhyp}{Hypothesis}
   
\stepcounter{hyp}

\begin{document}

\begin{frontmatter}

\title{Algorithmic Transparency in Forecasting Support Systems}

\author[a,*]{Leif Feddersen}
\ead{feddersen@bwl.uni-kiel.de}
\affiliation[a]{organization={Institute for Business, Christian-Albrechts-University Kiel},
                country={Germany}}
\affiliation[*]{Corresponding author}

\begin{abstract}
Most organizations adjust their statistical forecasts (e.g. on sales) manually.
Forecasting Support Systems (FSS) enable the related process of automated forecast generation and manual adjustments.
As the FSS user interface connects user and statistical algorithm, it
is an obvious lever for facilitating beneficial adjustments whilst discouraging harmful adjustments.
This paper reviews and organizes the literature on judgemental forecasting, forecast adjustments, and FSS design. 
I argue that algorithmic transparency may be a key factor towards better, integrative forecasting 
and test this assertion with three FSS designs that vary in their degrees of transparency 
based on time series decomposition.
I find transparency to reduce the variance and amount of harmful forecast adjustments.
Letting users adjust the algorithm's transparent components themselves, however, leads to widely varied and overall most detrimental adjustments.
Responses indicate a risk of overwhelming users with algorithmic transparency without adequate training.
Accordingly, self-reported satisfaction is highest with a non-transparent FSS. 
\end{abstract}


\end{frontmatter}

\begin{acronym}	
    \acro{AE}[$AE$]{Algorithm Effect}
    \acro{AF}[$DF$]{Adjustment Frequency}
    \acro{API}{Application Programming Interface}	
    \acro{AV}[$AV$]{Adjustment Volume}
    \acro{DSS}{Decision Support System}
    \acro{DV}[$DV$]{Deviation Volume}	
    \acro{ES}{Exponential Smoothing}
    \acro{FSS}{Forecasting Support System}
    \acro{GAM}{Generalized Additive Model}
    \acro{GUI}{Guided User Interface}
    \acro{HIT}{Human Intelligence Task}
    \acro{iid}[i.i.d.]{independent and identically distributed}
    \acro{IP}{Internet Protocol}
    \acro{MAE}[$MAE$]{Mean Absolute Error}
    \acro{MAPE}[$MAPE$]{Mean Absolute Percentage Error}
    \acro{mTurk}{Amazon Mechanical Turk}
    \acro{O}{Opaque FSS Design Treatment}
    \acro{rMAE}[$rMAE$]{Relative Mean Absolute Error}
    \acro{RNN}{Recurrent Neural Network}
    \acro{SES}{Simple Exponential Smoothing}
    \acro{SHF}{Simulated Historic Forecast}
    \acro{SNAP}{Supplement Nutrition Assistance Program}
    \acro{SOP}[S\&OP]{Sales and Operations Planning Process}
    \acro{T}{Transparent FSS Design Treatment}
    \acro{TA}{Transparently Adjustable FSS Design Treatment}
    \acro{USD}{United States Dollar}
\end{acronym}

\include{text/1-Introduction}
\include{text/2-LiteratureReview}
\include{text/3-Hypotheses}
\include{text/4-Background}
\include{text/5-Study}
\include{text/6-Conclusion}

\bibliography{bib/bibliography}
\bibliographystyle{apalike}
\include{text/x-appendix}

\end{document}

%% file: text/1-Introduction.tex
\section{Introduction}
\label{sec:intro}

Man or machine—who is the better decision-maker? This question, some would argue, was answered almost 70 years ago in favor of machines by \citet{meehl1954} in the context of clinical versus statistical disease prediction. Despite recent advances in artificial intelligence—making machines even more flexible yet precise decision-makers—humans often insist on having the last say. Past misadventures with decision automation—for instance, the automated demand forecasting issues at Nike \citep{Worthen2003}—and legitimate warnings about a potential 'intelligence explosion' in the future \citep{Russell2015} suggest that this hesitation is justifiable. However, relying solely on human decision-making brings with it a long list of documented flaws. Using their respective strengths, the reconciliation and integration of human and algorithmic decision-making appears to be a sensible overarching goal for research and practice.

Forecasting is a crucial organizational function, which supports the \acf{SOP}. The generation and adjustment of forecasts is often facilitated and supported by a \acf{FSS} \citep[p.233]{Boylan2010}, a subcategory of \acfp{DSS}. Its common understanding is that of software that connects to the organization's database, visualizes data, produces statistical forecasts, and allows for incorporating managerial judgment \citep[p.352]{Fildes2006a}.

Most organizations first automatically create quantitative forecasts, e.g., on sales or demand, and subsequently adjust them manually \citep{Fildes2009}, often via an \ac{FSS} user interface. Case studies on forecasting processes suggest that the effects of adjustments on forecasting accuracy are highly variable, often resulting in only a small net improvement at the expense of valuable working hours \citep{Fildes2021}. Controlled laboratory studies on judgmental forecasting and adjustments paint an even more pessimistic picture of human extrapolation capabilities. Human forecasters suffer from biases, noise\footnote{Biases are systematic deviations from rational choices, while noise refers to random variations in judgment that have received less scientific attention as of yet \citep{kahneman2021noise}} \citep{Hogarth1981}, and frequently unwarranted aversions to algorithmic advice \citep{Dietvorst2015}.

Mitigating these problems and facilitating more accurate forecasts requires understanding the forecasting process from the manager's perspective. For that purpose, I propose a simple task-oriented framework by which I will conceptualize the forecasting process and the related literature.

Its central actor is the human decision-maker. They generate, adjust, and sign off quantitative forecasts for a given entity. Doing so, they face four primary inputs, which, \emph{mediated by human behavior}, predict the main output: forecasting accuracy. Namely, these inputs are:

\begin{enumerate}
    \item The historical time series to be extrapolated
    \item Extra information relating to that time series
    \item The task environment dictating explicit and implicit goals
    \item The \ac{FSS} facilitating forecast interactions
\end{enumerate}

The inputs are not independent of each other. Some inputs dictate the potential scope of others or can be viewed in a hierarchical relation. For example, the scope of available information is highly influenced by the task environment, specifically the organization's information and knowledge management. In turn, the available information dictates what kind of algorithms are applicable. The type of algorithm (\ac{FSS} backend) influences how transparent the \ac{FSS} frontend can be. These considerations show that while the compartmentalized examination of the components at play is valuable and necessary, a complete and relevant understanding requires the subsequent integration considering all their interdependencies.

This work's empirical study focuses on the central role of the \ac{FSS}'s \acf{GUI}. Its facilitative role in forecast generation and adjustment and interaction with the other inputs make it an obvious lever for mitigating detrimental user inputs while encouraging beneficial ones. As noted by \citet{Goodwin2015}, commercial \ac{FSS} frequently lack a user-centric and behaviorally informed design. A behaviorally informed \ac{FSS} design reflects and extends the user's understanding of the decision object, their ``mental model'' \citep{odonnel2000}, and thereby promises more informed decision-making.

One aspect of such systems is \emph{algorithmic transparency}, ``the extent [to] which the inner workings or logic of the automated systems are known to human operators to assist their understanding about the system'' \citep[p.611]{Seong2008}. Algorithmic transparency is thought to promote user trust and thereby mitigate unnecessary adjustments \citep{Sheridan1988,Haubitz2021}. In the domain of time series forecasting, decomposing algorithms are the natural choice for a transparent representation because their components, e.g., seasonal effects, and their integration through addition or multiplication, are familiar, tangible concepts.

Therefore, the research presented here asks: \emph{To what extent can algorithmic transparency through decomposition facilitate more accurate forecasts by mitigating harmful adjustments whilst promoting beneficial adjustments?}

To start answering this question, I implement three \ac{FSS} designs with varying degrees of transparency. Transparency is achieved by dynamically displaying each forecast's constituent additive components: trend level, weekly cycle, yearly cycle, and events. The transparently adjustable design allows the individual adjustment of those components. I conduct an empirical study on \acf{mTurk} to examine the effects of transparency on adjustment volume, adjustment quality, and user satisfaction utilizing real-world Walmart sales data from the M5 competition \citep{m5comp}.

The following section presents an overview of the relevant literature, informing when, why, and how human judges adjust statistical forecasts. Additionally, findings on pure judgmental forecasting are included due to their close relation to adjustments. Based on the research gaps identified in the literature review, Section~\ref{hypotheses} derives this work's hypotheses. Section~\ref{background} lays out the algorithmic model and dataset used for the experiment. Section~\ref{mainstudy} details the experimental procedure and results, focusing on algorithmic transparency, and discusses the findings. Section~\ref{conclusions} concludes this work.

%% file: text/2-LiteratureReview.tex
\section{Literature Review}
\label{literaturereview}

This literature review begins by examining isolated forecasting tasks with minimal context provided to human judges, then extends to experiments and case studies where additional information is incorporated. These sections focus on human biases in judgment, rooted in the 'heuristics and biases' program \citep{Tversky1974}. Subsequently, the task environment is explored from the 'fast and frugal' perspective \citep{gigerenzer1991}, highlighting discrepancies between laboratory studies and practice. Finally, I discuss empirical findings on \ac{FSS} design aimed at mitigating these challenges, including studies that structure information input and output in ways transferable to \ac{FSS} design.

\subsection{Isolated Forecasting Tasks}
\label{isolatedforecasting}

In isolated forecasting tasks, where subjects are provided only with the historical time series or a statistical baseline forecast, they exhibit decision biases leading to inferior predictions compared to algorithmic forecasts \citep[p. 126]{Hogarth1981}. These biases have been established within the general 'heuristics and biases' program \citep{Tversky1974} and adapted for judgmental forecasting, such as the anchor and adjust heuristic, while others are forecasting-specific\footnote{It would be erroneous to view heuristics and biases as flawed essences of human decision-making. Instead, they are task-dependent behavioral manifestations of human cognition. Thus, several biases can be caused by the same cognitive process \citep{Hilbert2012}, and different cognitive processes can manifest in the same observed bias \citep{Edelman2001}.}

\paragraph{Anchor and Adjust} The anchor and adjust heuristic suggests that human judges initially form or possess a fixed value before engaging deeply in a decision task. In time series forecasting, this anchor may be the most recent observation \citep{Kremer2011}, the series' mean \citep{Bolton2012}, an initial prediction \citep{Lim1996}, or another salient cue. Decision-makers often adjust conservatively from this anchor, regardless of its rational predictive strength. Interestingly, \citet{Lawrence1995} found that judges excessively adjusted the supposed anchor—a normative statistical prediction—which may be explained by algorithm aversion, discussed later.

\paragraph{Noise Modeling and Hindsight Bias} Noise modeling refers to the tendency of human judges to perceive patterns in random fluctuations and extrapolate them into the future, leading to inaccurate forecasts. For instance, \citet{Connor1993} challenge the assumption that humans adapt better to changing environments than algorithms. Their study shows that MBA and Ph.D. students forecasting artificial time series with level changes are not only slower than exponential smoothing methods in detecting these changes but also attempt to extrapolate from random noise. Similarly, \citet{Kremer2011} describe the overweighing of recent observations relative to the complete history—a phenomenon they term system neglect—which leads to overreaction in stable environments and underreaction in unstable ones. This behavior is closely related to the base-rate fallacy \citep{Tversky1974}, where judges ignore underlying probabilities when faced with salient cues.

An underlying cause of noise modeling is hindsight bias, where decision-makers perceive past events as more predictable than they were \citep{roese2012}. This bias leads them to overlook alternative outcomes, fostering overconfidence in their predictive abilities. The combination of perceiving illusory patterns in randomness and overestimating one's capabilities exacerbates errors in judgmental forecasting.

\paragraph{Extreme Preference} Noise modeling can lead to extreme forecasts, and evidence suggests that extremeness is a desirable attribute in judgmental forecasts and adjustments. For instance, \citet{kahneman1973psychology} found that students given less predictive information did not produce more conservative forecasts as rationality would suggest; instead, they maintained extreme predictions. Similarly, \citet{thaler1990} analyzed real-world security analysts' earnings forecasts and observed that their adjustments resulted in forecasts that were, on average, 54\% more extreme for a one-year horizon and 117\% for a two-year horizon. \citet{dietvorst2020people} experimentally confirmed that decision-makers exhibit diminishing sensitivity to errors, preferring high-variance human judgments over low-variance algorithmic predictions, even when the latter are more accurate. This preference for high-variance models is also observed in a case study by \citet[p.12]{Fildes2021}, where some managers are described as intolerant of noise (i.e., unexplained residuals and random error) in fitted models.

\paragraph{Trend Dampening} Another observed behavior is trend dampening, where judges reduce the magnitude of linear trends (i.e., regress to the mean), especially for downward-sloping series \citep[e.g.,][]{lawrence1989,Webby1994}. As demonstrated by \citet{Gardner1985}, this behavior often reflects the rational expectation that most trends do not continue indefinitely. While humans are notoriously bad at grasping exponential growth, the concept of diminishing returns (i.e., logarithmic or damped linear growth) is familiar to most people.

\paragraph{Gambler's Fallacy} The gambler's fallacy is a robust phenomenon where decision-makers expect a reversal after a streak of similar outcomes. In forecasting, its manifestation is less straightforward. \citet{Kremer2011} operationalize forecasts that go against the recent pseudo-trend of a stationary time series as a result of the gambler's fallacy but find that only 11\% of subjects forecast accordingly. Additionally, \citet{Petropoulos2016} identify several detrimental forecast adjustments attributable to other decision biases but not the gambler's fallacy. These findings suggest that the gambler's fallacy is not prevalent in typical forecasting tasks unless specific cues lead judges to erroneously anticipate mean reversion.

\subsection{Forecasting with Extra Information}
A common objection against algorithmic predictions is, "I know things the model cannot know." Indeed, omitted variables can lead to model deficiencies. Paul Meehl famously coined the "broken-leg" cue example, describing a statistical model predicting cinema attendance. Regardless of the model's average accuracy, knowing a person has broken their leg allows us to predict with certainty they will not attend the cinema \citep{meehl1954}. In forecasting, a "broken-leg" cue might be knowledge of a supply shortage, a recent trend change, or a planned promotion. Importantly, by definition, the cue must not be included in the model; otherwise, correcting for it leads to the double counting bias \citep{Jon}. I now review findings on how well human judges incorporate extra-model information.

\subsubsection{Laboratory Research}

\citet{Lim1996a} employ a simple experimental setup: Participants forecast soft-drink sales at Bondi Beach given historical sales. They are then provided either with a statistical forecast, temperature data (either highly predictive or mildly predictive), or both, and can adjust their initial forecast. Results show that participants are sensitive to the temperature data's predictive validity and utilize the statistical forecast. However, they do so suboptimally and do not increase their weighting of the extra information over time. Interestingly, they prefer the statistical model over the temperature data, possibly because temperature uses a different scale.

A study by \citet{Harvey1994} uses a more complex forecasting problem. Participants are asked to forecast the sinusoidal pattern of train passengers and criminals on the train. While subjects are relatively precise in separate forecasts, their performance drops significantly when forecasting passengers given only the criminals' time series. Correct forecasts would have required participants to internalize the relationship between the two series, which involved a phase shift and amplitude difference.

Additional studies by \citet{garner1982} and \citet{becker2007,becker2008} yield similar results, indicating that humans incorporate quantitative extra information suboptimally \citep[p.462]{Leitner2011}, especially if the relationship is nonlinear. This finding is not surprising given human limitations in processing quantitative data accurately; an adequate statistical model would easily produce optimal forecasts in these conditions.

Soft information refers to non-quantifiable cues, such as rumors, gossip, or advice. Operationalizing soft information in laboratory research is challenging because providing it explicitly makes it less 'soft' in the traditional sense \citep{Mintzberg1975}. With that in mind, several studies have supplied human forecasters with soft information.

In \citet{remus1998}, business students forecast time series given correct, incorrect, or no soft information framed as rumors about future movements. Participants successfully utilize correct information and quickly identify incorrect information. Receiving correct information after incorrect information leads to short-term discounting of new information, but this effect does not persist long-term.

\citet{Goodwin2011} explore the effectiveness of restricting and guiding \ac{FSS} designs. They find that participants are particularly persistent in acting on rumors, which only indicate the direction (not magnitude) of necessary adjustments and are correct only 62.5\% of the time. They hypothesize that rumors were more salient in their study than other information sources and, in practice, are often associated with obligations within the organizational hierarchy.

\subsubsection{Case and Field Studies}
\label{subsec:casestudies}
Case studies provide a better gauge for the effects of managerial intervention on forecast accuracy but, lacking experimental controls, make statements about cause and effect more challenging. With these qualifications in mind, we discuss relevant findings.

\citet{Blattberg1990} examine expert and statistical forecasts in five companies. Their analysis reveals that both forecast types have comparable cross-validated $R^2$ and that combining model and expert forecasts improves accuracy. They calculate that managers' forecasts explain, on average, 25\% of the model's unexplained variance, supporting the notion that managerial intuition is valuable but flawed.

\citet{Sanders1992} compare real-world forecasts of experienced warehouse planners with those of students and a statistical ensemble method. Their results show that the statistical forecast performs similarly to experts for time series with low variability, but experts have an advantage for highly variable series.

Similarly, \citet{Edmundson1988} distinguish practitioners with product and industry knowledge from those with only industry knowledge. Their results show that only product knowledge leads to significantly more accurate forecasts.

A series of studies by \citet{Mathews1986,Mathews1989,Mathews1990,Mathews1992} involving a UK healthcare company found that judgmental adjustments increase forecasting accuracy overall but introduce an optimism bias. Managers are effective at selecting inaccurate statistical forecasts for revision, and large adjustments more often lead to improvements than small adjustments.

\citet{Eroglu2010} analyze the influence of personality and motivational factors on forecasting biases. Notably, the propensity to adjust forecasts is reliably determined by work experience in the current position, an internal locus of control, and a challenge-seeking orientation.

\citet{Fildes2021} provide a detailed case study of the forecasting process in a UK pharmaceutical company. They find that while large negative adjustments added value, positive adjustments, especially small ones, decreased accuracy on average. The managers displayed typical noise modeling behavior and adjusted forecasts to satisfy colleagues' desires for smooth-looking but overfitted models.

\subsection{The Task Environment}

\subsubsection{Ecological Rationality}
\label{ecorationality}
The previous sections paint a pessimistic picture of human involvement in forecast creation. However, some case studies find that managers enhance statistical forecasts under certain circumstances. These observations are consistent with the perspective of ecological rationality proposed by \citet{Luan2019}, which offers an alternative to the "heuristics and biases" program by \citet{Tversky1974}.

Both approaches agree that humans often use heuristics—simple decision rules that ignore some information and do not involve mathematical optimization. They differ in interpretation: While heuristics can lead to biased decisions in inappropriate environments, they can produce "fast and frugal" decisions when matched to the environment \citep[p.2]{Luan2019}. The appropriateness of a heuristic for a given environment determines its accuracy, and selecting an appropriate heuristic is a hallmark of an intelligent decision-maker \citep[p.178]{gigerenzer2015simply}.

Therefore, successful practitioners may be examples of effectively applied ecologically rational heuristics. Heuristics are often better suited for real-world situations characterized by uncertainty, whereas statistical models are more appropriate for problems involving risk with known probabilities.

Ecological rationality also implies that forecast adjustments may reflect goals beyond improving accuracy. Personal and organizational circumstances can impose additional motives, leading to adjustments that seem irrational from an accuracy standpoint.

\subsubsection{The Organizational Environment}
\label{orgaenv}

The organizational environment can impose various motives for over-adjusting forecasts. These motives can arise from the adjuster's role within the organization or from the firm's functional and structural organization.

\paragraph{Individual-Driven Motives} Judgmental adjustments can serve as a signal that a manager has reviewed and approved the forecasts. Beyond this basic signal, adjustments can convey competence and credibility, especially in environments with low levels of post-adjustment accuracy tracking \citep{fildes2015}. Adjustments may also help managers maintain and improve their forecasting skills in case the algorithmic system fails \citep{bainbridge1983ironies,DEBAETS2018}.

Involvement in forecast creation increases ownership and acceptance of the forecast \citep[e.g.,][]{Lawrence2002}. This increased acceptance extends beyond the adjuster; for example, analysts may adjust forecasts to satisfy colleagues' preferences \citep[p.12]{Fildes2021}. \citet{Onkal2008} demonstrate experimentally that judges prefer judgmentally adjusted forecasts over pure statistical ones, especially when accompanied by a human-generated explanation.

\paragraph{Structure-Driven Motives}
In a functionally distributed organization, forecasts can have different implications for each functional area, leading to incentives to bias forecasts. \citet{mello2009impact} and \citet[p.143]{olivia2009} describe scenarios such as enforcing, filtering, hedging, sandbagging, spinning, second-guessing, and withholding, where forecasts are deliberately adjusted upwards or downwards due to various organizational motives.

\subsubsection{The Experimental Environment}
\label{expenv}

Laboratory settings can also facilitate unintended ecological rationality among subjects. "Winner takes all" incentives, as implemented in some studies \citep{Prahl2017,Lim1995,Lim1996}, nudge subjects to provide more extreme forecasts. This occurs because extreme deviations may increase the chance of outperforming competitors or the provided algorithmic forecast.

Incentivizing participants based on absolute performance is a potential remedy. However, defining an appropriate incentive scheme must account for loss aversion \citep{kahneman1979}. An effective incentive scheme for judgmental forecasting studies should encourage accuracy without promoting extreme adjustments.

\subsection{The Forecasting Support System}

An optimal \ac{FSS}, as formulated by \citet[p.354]{Fildes2006a}, "(i) improves the forecaster's ability to realize when judgmental intervention is appropriate and (ii) enables the user to apply accurate judgmental interventions when these are appropriate."

Design approaches toward these goals have been the subject of diverse empirical and theoretical contributions. Key areas include restrictiveness and guidance, model selection and parameterization, trust in algorithmic advice, and the structuring of informational inputs and outputs.

\subsubsection{Restrictiveness and Guidance}
\label{subsec:restrictivenessandguidance}

In his seminal work on \ac{DSS} design, \citet{silver1991} introduces the concepts of guidance and restrictiveness as "meta-support" for using a \ac{DSS}. For example, an \ac{FSS} could be highly restrictive, allowing no human input, or it could permit adjustments and guide users on when and how to adjust forecasts.

\paragraph{Effort Manipulations} Subtle restrictiveness can be implemented by making desirable forecasting strategies less effortful than less desirable ones \citep[p.355]{Fildes2006a}. Based on the effort-accuracy framework by \citet{payne1993}, this approach suggests that increasing the effort required to make adjustments can reduce harmful adjustments. \citet{Goodwin2000} find that requiring explicit requests for adjustments or asking for reasons reduces harmful adjustments while maintaining beneficial ones.

\paragraph{Explicit Support} \citet{Goodwin2011} compare the effects of restrictive and guiding \ac{FSS} designs. They find that guidance can encourage necessary adjustments and discourage unnecessary ones, but overly restrictive designs may backfire, leading to larger and more harmful adjustments.

\subsubsection{Model Selection and Parameterization}
\label{modelselection}
Allowing users control over model selection and parameterization can increase satisfaction and reduce subsequent adjustments \citep{Lawrence2002,Petropoulos2018}. While human decision-makers may not always select the optimal model, they are often effective at avoiding the worst options. This aligns with case study reports where managers successfully identified and corrected poor statistical forecasts.

\subsubsection{Algorithm Aversion and Trust}
\label{subsec:algoaversion}

Despite the demonstrated superiority of algorithmic predictions in many contexts \citep{meehl1954,Dawes1979TheRB}, algorithmic advice remains underutilized. \citet{Dietvorst2015} coin the term "algorithm aversion" to describe this tendency.

A systematic review by \citet{Burton2019} identifies five major drivers of algorithm aversion: false expectations, lack of perceived control, cognitive compatibility, lack of incentives, and divergent rationalities. These factors contribute to underutilization of algorithmic advice, especially after observing algorithm errors.

Experimental research shows that judges rely less on algorithmic advice after seeing it err, even when it outperforms human judgment \citep{Prahl2017,Dietvorst2015}. This may be due to inflated initial expectations of algorithmic accuracy and skepticism about the algorithm's ability to learn from mistakes \citep{berger2021watch}.

\subsubsection{Algorithmic Transparency} 
Algorithmic transparency—the extent to which the inner workings or logic of automated systems are known to human operators—can mitigate algorithm aversion \citep[p.611]{Seong2008}.

\paragraph{Understanding the Algorithm}
Transparency enables users to develop a mental model of the algorithm, facilitating trust and appropriate reliance. However, making complex algorithms transparent requires higher levels of abstraction. Additionally, if an algorithm is perceived as too simple upon being made transparent, users may distrust it, leading to disuse \citep{Haubitz2021}.

\paragraph{Understanding the Object}
Transparent algorithms can also help users understand the decision object better. By tracking the decision process, users can learn about relationships in the data and intervene more precisely when algorithmic decisions contradict extra-model information.

While algorithmic transparency holds promise, research in the context of time series forecasting is limited. Further exploration is needed to understand its potential benefits and pitfalls.

\subsubsection{Structuring and Decomposition}
\label{decomp}
Structuring judgmental decisions through decomposition is a well-established method to reduce bias and noise. Experiments have shown that decomposing time series into components like trend and seasonality allows human forecasters to make more accurate predictions.

\citet{Edmundson1990} developed an application called GRAFFECT, guiding users to identify trends and seasonal patterns before forecasting. This structured approach resulted in forecasts significantly better than unstructured judgmental forecasts and comparable to statistical methods.

\citet{Marmier2010} focus on integrating expert knowledge about event effects into forecasts. By categorizing factors (e.g., transient factors, trend changes) and incorporating them structurally, they demonstrate significant improvements over purely statistical forecasts.

\citet{Asimakopoulos2009} compare user satisfaction among \ac{FSS} prototypes with varying levels of information structuring. Participants preferred highly structured information and valued opportunities to document and reflect on their reasoning.

\subsection{Visual Design and Elements}
\label{subsec:visualdesign}

The visual display of information in an \ac{FSS} can significantly impact user interaction. The format used to display data, the arrangement of elements, and the prominence of information all influence decision-making \citep{silver1991}.

For example, \citet{Harvey1996} find that graphical displays help subjects identify trends but may also lead to perceiving trends in noise. \citet{Theocharis2018} show that line graphs prime viewers to see data as sequences, which can be beneficial for trended data but may encourage erroneous trend identification in random data.

Bar graphs, while effective at conveying metrics, can introduce biases. \citet{newman2012} demonstrate a "within-the-bar" bias, where subjects perceive points within the bar as more likely. \citet{harvey2012} show that bar graphs lead to systematically lower forecasts than points and lines for trended series.

\subsection{Literature Review Summary}
This literature review highlights the current state of judgmental forecasts and adjustments, the handling of extra information, the role of the task environment, and the central importance of \ac{FSS} design. Human judges are generally less accurate at extrapolating univariate time series than statistical methods and often struggle to incorporate extra-model information effectively. Experienced forecasters in practice can enhance statistical forecasts, but results are mixed, and the cost of human intervention must be considered. Adjustment motives may extend beyond improving accuracy due to personal and organizational factors.

The critical role of the \ac{FSS} has been widely recognized. Research shows that users appreciate transparent explanations, structure, and opportunities to incorporate their judgment. Attempts to restrict or guide \ac{FSS} usage are promising but can backfire if not carefully designed. Algorithmic transparency is underexplored in the \ac{FSS} context and presents potential for fruitful research, connecting well with successful concepts like forecast explanations and structured decision-making.

%% file: text/3-Hypotheses.tex
\section{Derivation of Hypotheses}
\label{hypotheses}.

The literature review revealed that algorithmic transparency and related concepts like understandability and explanations positively affect advice utilization. However, to the best of my knowledge, no experimental study to date has examined algorithmic transparency in the context of an \ac{FSS}-aided forecasting task.

I posit that algorithmic transparency should improve all relevant metrics in a scenario where a sales manager reviews, adjusts, and signs off algorithmic forecasts.
Overall, transparency should increase trust in the statistical forecast and lessen the desire to adjust it:

\begin{subhyp}\label{hyp:1a}
    Increasing \ac{FSS} transparency leads to a smaller adjustment volume.
\end{subhyp}

Since most adjustments, especially without extra predictive information, harm accuracy, it follows that:

\begin{subhyp}\label{hyp:1b}
    Increasing \ac{FSS} transparency leads to higher forecast quality post adjustments.
\end{subhyp}

The first two hypotheses relate to the quality of final forecasts and adjustment behavior. For an \ac{FSS} to be beneficial, it is also necessary that it is acceptable to its users; otherwise, it might go unused in the long run. Transparency is a promising feature to increase acceptance as it can foster understanding and learning about the model and the modeled system. Users might even enjoy exploring how an algorithm interprets a time series. Thus, Hypothesis~\ref{hyp:2} posits:

\begin{hyp}\label{hyp:2}
    Increasing \ac{FSS} transparency leads to higher user satisfaction.
\end{hyp}

The next section describes the background of the experimental study aimed at exploring these hypotheses.

%% file: text/4-Background.tex
\section{Experimental Background}
\label{background}

\subsection{Prophet}
\label{prophet}

An integral part of an \ac{FSS} is the underlying model generating forecasts. For algorithmic transparency to be effective, the model must be decomposable into interpretable parts. Prophet accommodates these requirements and is therefore used to generate all forecasts for this work’s experiment \citep{taylor2018}. Below, I briefly outline its core components and practical implications.

Prophet was developed to allow robust, automated forecasting at scale while enabling flexible integration of domain knowledge by analysts without deep time series expertise. It is a \acf{GAM} with the following formulation:
$$
y(t)=g(t)+s(t)+h(t)+\epsilon_{t}
$$
where $y(t)$ represents the forecast, $g(t)$ is the non-periodic piecewise trend, $s(t)$ the periodic seasonalities, $h(t)$ external regressors like holidays or events, and $\epsilon_{t}$ is the error term.

Prophet’s trend component is piece-wise between changepoints, which allows for flexibility in adapting to trend changes. Seasonal cycles can be modeled with an arbitrary number of Fourier series, and external regressors like holidays can be included. Prophet’s backend, built in Stan, provides posterior estimates for all parameters, offering probabilistic forecasts.

The main advantage of Prophet is its interpretability and its suitability for the "Analyst in the Loop" process, where flagged forecasts can be reviewed and adjusted by analysts based on their domain knowledge. Prophet flags forecasts based on criteria like underperformance against benchmarks (e.g., naive forecasts) or specific performance thresholds (e.g., maximum \ac{MAPE}). Adjustments can be made in either a programming environment (Python/R) or an interactive application (e.g., Voila or Shiny). 

\subsection{M5 Competition Data}
\label{mcompetitions}

The M5 Competition, launched in 2020, provided Walmart sales data for forecasting. It includes daily sales data for 3,049 products across 10 stores over a 1,941-day period, with a total of 42,840 hierarchical time series. The dataset includes extra features like holidays, promotions, and prices, allowing for complex, real-world forecasting scenarios \citep{makridakis2022m5}.

This dataset was selected for its relevance to retail forecasting, offering both time series data and rich external information. Intermittent demand, which is common in business time series, presents a particular modeling challenge addressed by Prophet.

\subsection{Data Selection and Model Specifications}
\label{sec:dataselection}

For the experiment, I select a subset of the M5 dataset, specifically focusing on "foods" to ensure relevance and reduce cognitive load in the experiment. Ten time series were chosen for their comparability, with a forecast horizon of 14 days to reflect realistic decision-making processes while simplifying participant tasks. Prophet models were optimized based on cross-validation, primarily tuning the trend and seasonality regularization parameters.

\subsection{Participant Selection and Recruitment}
\label{participants}

Participants are recruited from \ac{mTurk}, a platform allowing fast and cost-effective collection of data. Using cloudresearch \citep{litman2017turkprime}, participant selection is enhanced by filtering for only US-based Workers with management or sales roles, or students aiming for such positions, to increase external validity.
I conduct pre-studies to screen for Workers showing engagement with forecasting tasks. Those were invited to participate in the main experiment. 

\subsection{Incentive Scheme}
\label{incentivescheme}

Performance-dependent incentives are critical in ensuring truthful and engaged responses. In this study, I incentivize participants based on improvements over algorithmic forecasts. Each adjustment is judged on its ability to improve forecast accuracy, encouraging careful consideration without rewarding extreme adjustments. This scheme is designed to reflect organizational environments where adjustments are often not tracked, allowing for selective reporting of beneficial adjustments.

\subsection{Measured Indicators}
\label{indicators}

To test the hypotheses, several indicators are measured:

\paragraph{Adjustment Volume} \acf{AV} measures the sum of absolute adjustments made by participants relative to optimal adjustments for 100\% accuracy:

\begin{equation}\label{eq:volume}
\ac{AV} = \frac{\sum_t^T{|y^{model} - y^{final}|}}{\sum_t^T{|y^{model} - y^{truth}|}}
\end{equation}

\paragraph{Adjustment Frequency} \acf{AF} captures how often participants adjust forecasts, calculated as the fraction of forecasts that were modified:

\begin{equation}\label{eq:adjfreq}
Adjustment \: Frequency_p= \frac{\sum_n^N{I(AV)}}{N}
\end{equation}

\paragraph{Relative MAE} \acf{rMAE} is the ratio of the \acf{MAE} of the participant’s forecast to the algorithmic forecast:

\begin{equation}\label{eq:rmae}
rMAE_{a,b} = \frac{MAE_a}{MAE_b}
\end{equation}

\paragraph{Mean Absolute Percentage Error (MAPE)} measures the average percentage error between the participant’s forecast and the actual values:

\begin{equation}\label{eq:mape}
MAPE = \frac{1}{T}\sum_t^T{\frac{|y_t - \hat{y}_t|}{y_t}}
\end{equation}

%% file: text/5-Study.tex
\section{Experimental Study}\label{mainstudy}

This work's experimental study examines the effects of algorithmic transparency on user interaction and satisfaction with an \ac{FSS}.
To that end, I implement three \ac{FSS} designs with increasing degrees of transparency and let participants review, adjust, and sign-off
Prophet forecasts within the respective interfaces. I hypothesized that algorithmic transparency would lead to
overall less but better adjustments and increase user satisfaction.

\subsection{Operationalizing Algorithmic Transparency}

I operationalize \ac{FSS} transparency by juxtaposing three \ac{FSS} designs:
The \acf{O} makes no attempts at explaining the underlying algorithmic model.
The \acf{T} makes the model transparent to the user by dynamically displaying its components.
The \acf{TA} further allows the individual adjustment of the components.\footnote{Throughout this paper, the acronyms O, T, and TA refer to the respective treatments, participants in those treatments, the FSS designs, and the associated adjectives. The intended meaning will be clear from the context.}

\paragraph{Transparently Adjustable}

The \ac{TA} \ac{FSS} operationalizes the concept of algorithmic transparency
to the maximum extent, that I view behaviorally feasible in the context
of an online experiment. 
\citet[p.513]{Bunn1991} conclude their literature review on the integration
of judgmental and statistical forecasting with the call for
``an interactive decomposition structure'', explicitly referencing the
judgmental seasonal decomposition approach by
\citet{Edmundson1990} as a promising basis towards a synthesis
of judgment and algorithm. The \ac{TA} \ac{FSS} aims to be exactly that. 
It allows for judgmental overrides, just like GRAFFECT \citep{Edmundson1990},
but a statistical algorithm sets all initial parameters (see Section \ref{sec:dataselection})

\begin{figure}
	\includegraphics[width=1.0\textwidth]{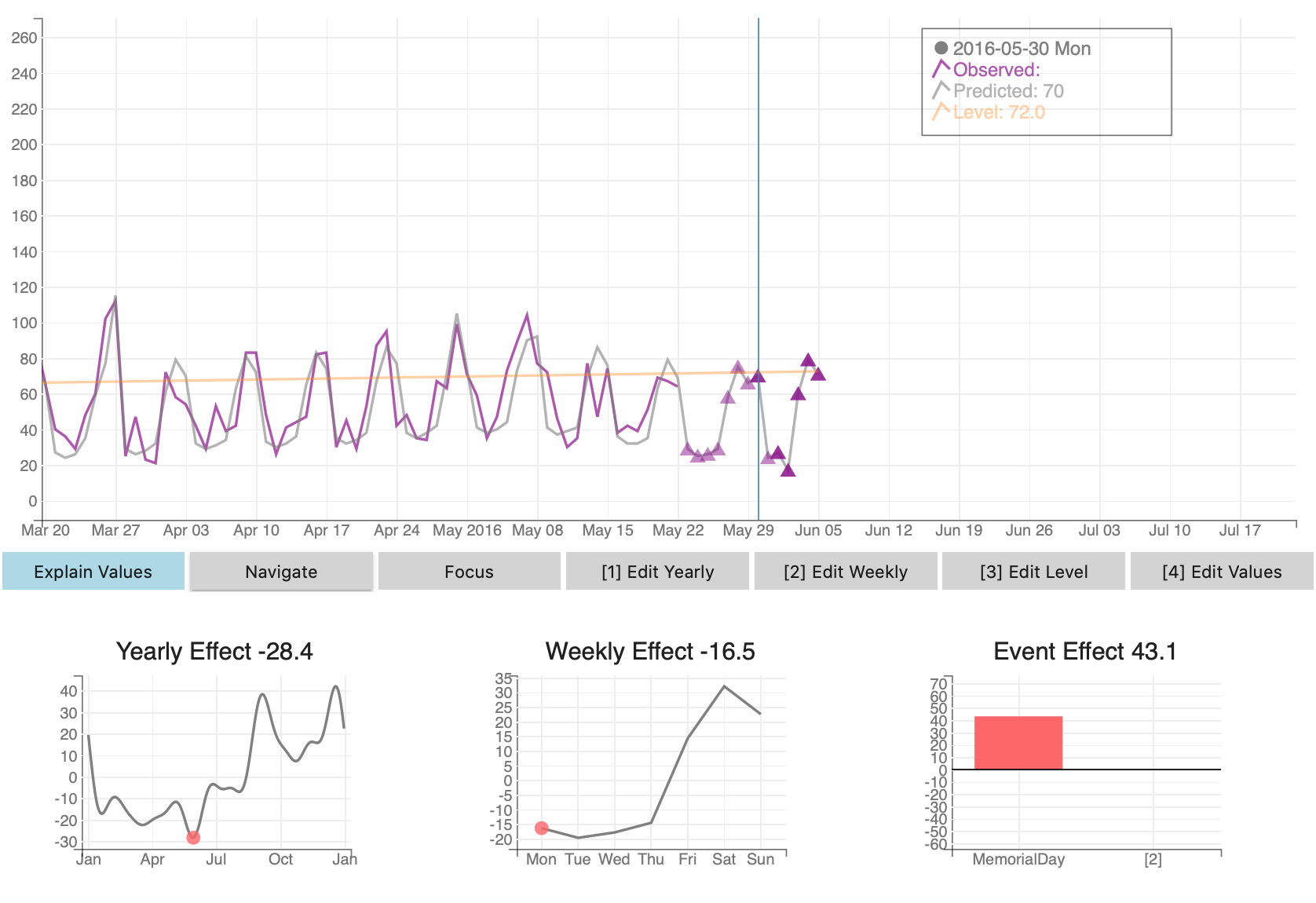}
	\caption{Transparently Adjustable FSS design in its final iteration.}
	\label{fig:study3TA}
\end{figure}

Figure \ref{fig:study3TA} showcases a typical screenshot when using
the \ac{TA} \ac{FSS}. 
The user has selected the function ``Explain Values'', which activates a
vertical selector bar. By positioning the selector bar on the
May 30, 2016, the graph updates its legend to indicate a predicted value
of 70 and a level of 72. As participants have learned, the prediction is the sum of the level and the three effects depicted in the three smaller graphs.

The bottom left graph shows the effect size of the yearly seasonality,
numerically displayed in the title, the overall pattern of yearly seasonality,
and the current position in that cycle. In the example, one observes
that the end of May is associated with particularly low sales, so Prophet
subtracts 28.4 sales units. Analogously, the bottom center graph shows
that Mondays are associated with a 16.5 decrease in sales.
The bottom right graph changes its x-axis labels according to the
the events that day (up to two for the M5 dataset) and displays
the sum of these events in its title. In the example, Prophet adds 43.1 sales units due the effect of \emph{Memorial Day}.
More details on \ac{FSS} functionality is integrated in the next Section \ref{sec:study3procedure}.

\paragraph{Transparent}
The \acf{T} is similar to \ac{TA}
but lacks the possibility to adjust the model's components individually.
Hence, adjustments are only possible by dragging the individual forecast values. 

\paragraph{Opaque}
The \acf{O} drops any explanations of the model forecast.
Hence, no trend line nor the effects graphs are displayed in that design. 
The ``Explain Values'' is therefore renamed to ``View Details'' as the selector bar
now only allows to extract details about date, weekday, fitted value, and observed value.

\subsection{Experimental Procedure}
\label{sec:study3procedure}

I set up the recruitment of 240 Workers on mTurk with the requirements described in Section \ref{participants}.

After entering their Worker ID, participants are informed
that the graph shows the sales data for
one out of three foods product. As a product manager,
they are responsible for the 14-day ahead forecasts for three such products and an algorithmic model aids them. 
The incentive is prominently displayed and reads that accuracy improvements over the model forecast through adjustments result in a bonus of 0.20 \acs{USD} per percentage point, up to 1 \acs{USD} per product, hence up to 3 \acs{USD} per \ac{HIT}. 

Next, for \ac{TA} and \ac{T}, participants learn that
the model understands the sales data ``step by step''.
First, the trend level (``general level of sales'')
is described and visualized in the graph with a gray line. 
In the next step, that same gray line morphs into
the sum of trend and seasonality. Participants learn about
the ``yearly pattern'', and the small graph for yearly effects
appears below the main graph. 
In the next step, weekly effects and the weekly graph are added analogously.
Finally, the model adds the event effects to the gray line to represent the ``model's final estimation''.

For the Opaque \ac{FSS} design, the steps mentioned above reduce
to the direct display of the fitted model without further explanations. 

Next, participants are asked to familiarize themselves with the functions for navigating the graph and exploring
the data:
\begin{enumerate}
	\item \emph{Explain Values} (\ac{TA} and \ac{T}) / \emph{View Details} (\ac{O}) - activate the selector bar to view further details about fitted and observed values.
	\item \emph{Navigate} - activates pan and zoom functions to navigate the data via mouse and scrollwheel. 
	\item \emph{Focus} - quickly focusses the graph around the forecast period
\end{enumerate}

Subsequently, participants explore the adjustment functions. 
Only participants in the \ac{TA} condition have the following options:

\begin{enumerate}
\item \emph{Edit Yearly} -  Switches the data content of the large main graph to display the yearly effects pattern plus the residuals it models.
The user can redraw the yearly effects and reset them. 
\item \emph{Edit Weekly} - Switches the data content of the large main graph to display the weekly effects pattern plus the residuals it models.
The user can adjust the weekly effects by dragging the white handle points and reset them. 
\end{enumerate}
A slider allows limiting the number of past fluctuations shown in \emph{Edit Yearly} and \emph{Edit Weekly}.
Figure \ref{fig:FSS3Weekly} shows a screenshot after enabling \emph{Edit Weekly}.
Only the past 38 weeks of fluctuations, as purple points, are shown.
Moving the slider limits that amount. Fluctuations are ordered per day by date (old to new, left to right), and their y-coordinates indicate
the respective residuals (model prediction without weekly effect) on that day.
After switching back to the main graph, there is a slight delay of 800ms until the main graph is smoothly updated
to allow for the visual perception of the change created through the adjustments of these components.
The last \ac{TA}-specific function is:

\begin{figure}
	\includegraphics[width=1.0\textwidth]{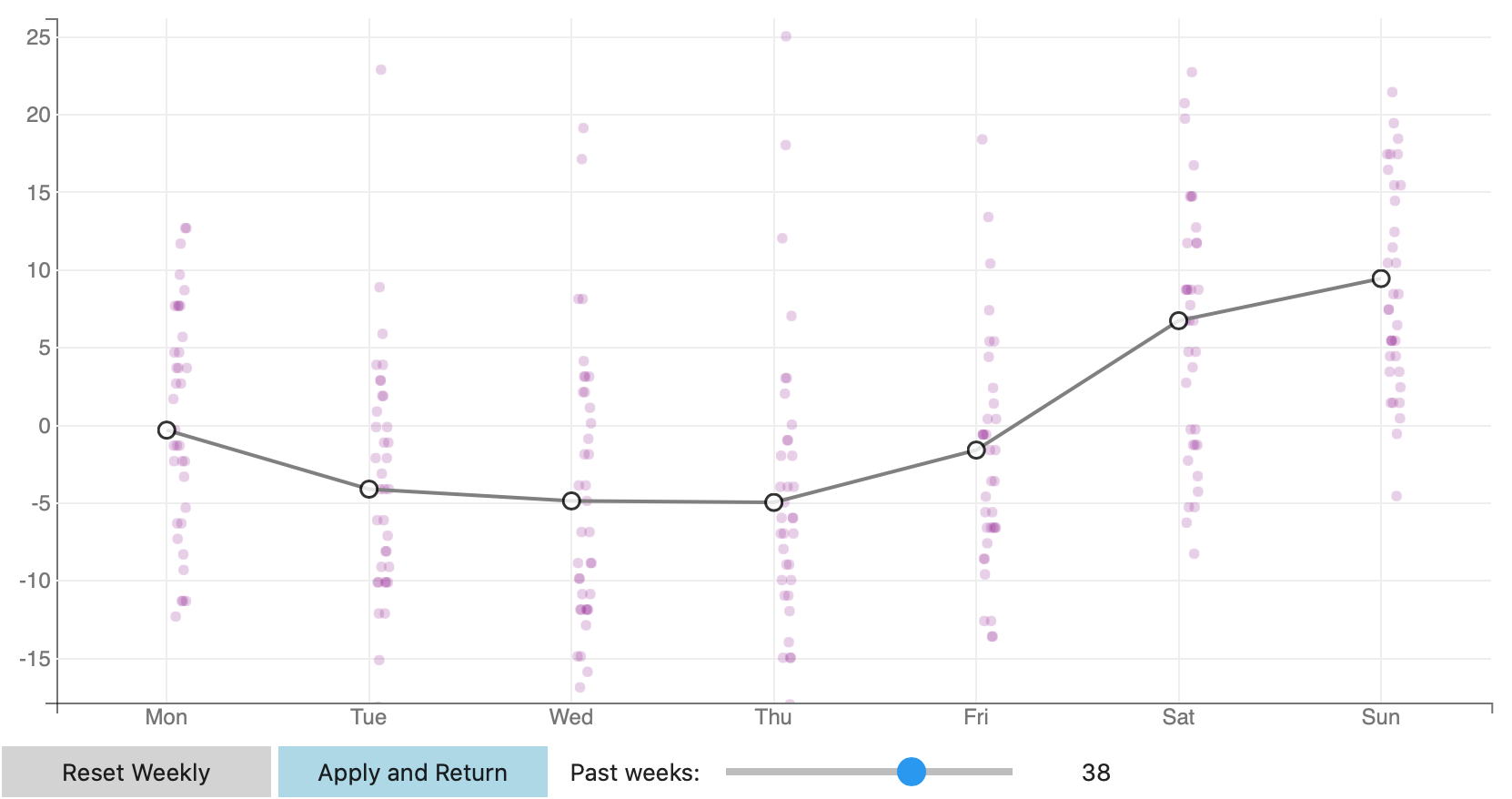}
	\caption{TA FSS: Detail view of weekly effects which allows for adjustments by dragging the white handle points.}
	\label{fig:FSS3Weekly}
\end{figure}

\begin{enumerate}
\item \emph{Edit Level} - Enables the trend line to be redrawn in the main graph. The forecast line is directly updated accordingly.
\end{enumerate}
Each adjusted component in \ac{TA} has an individual reset button in addition to the overall reset button. 
All \ac{FSS} designs allow the adjustment of single values:
\begin{enumerate}
\item \emph{Edit Values} - Enables the forecasted values to be adjusted via drag and drop. 
\end{enumerate}

Once participants feel familiar with the FSS functions, they are asked to click ``Start Signing Off''.
The instructions text reminds participants of the available functions and the bonus structure.
Participants can ``Sign Off'' a forecast with a button. If, however, they click ``Sign Off'' after less than
ten seconds of viewing the forecast, the button turns gray and asks ``So fast?'' for one second. 
Once all forecasts are signed off, participants indicate their approval of five statements
with a slider widget. These are:
\begin{enumerate}
\item I understand how the sales model creates its forecasts.
\item Such a forecasting app would be useful for product managers. 
\item I could bring my intuition into the forecasts.
\item I am satisfied with my signed-off forecasts. 
\item My adjustments were mostly motivated by the chance for a bonus. 
\end{enumerate}
A free-text field that can be left blank asks if one particularly liked or disliked something.
After confirming the answers, participants receive their secret key, accuracy measures, and earned bonus. 

\subsection{Overhaul in Response to Pre-Test}
A pre-test informed the previously outlined \ac{FSS} designs and experimental procedure. It implemented an early version of the experimental design, which revealed several design flaws. For the pre-test, I invited a small sample of ten Workers who displayed high levels of engagement in the pre-studies (as measured by engagement time and number of interactions), 
to partake in the planned experiment while commenting on things they liked, disliked, found confusing, or suspected to be confusing to others.
The comments turned out to be longer and more detailed than I had hoped but indicated that roughly one-half
of respondents was at some point confused about one or more crucial elements. 
Their points of confusion or criticism matched those of friends and peers who engaged with
my early \ac{FSS} prototype. I will present the lessons learned in terms of their implications for the final \ac{FSS} and experimental design,
contrasting the old and revised new versions:

\paragraph{Constant Task Display} \textit{Old}: An interactive tutorial was followed by a relatively
free interaction with the data and forecasts, meaning that there was no explicit task displayed
after the tutorial, and participants could move between product forecasts via a dropdown. 
Via a ``Help'' button, participants could request detailed descriptions of the 
task, incentive, and functions of the \ac{FSS}. Some respondents indicated that they did not know
their ``next step'' or whether the task had already started.
\textit{New}: Acknowledging that the free interaction, while resembling a realistic setting more closely,
is too overwhelming for an online experiment, the revised experiment always displays a text with the current task or next step (e.g., an explanation of how to adjust a component once selected) in the prominent top-left corner of the screen.
It is also unidirectional in that participants sequentially sign off forecasts and can not freely move between
product forecasts via a dropdown. 

\paragraph{User Interface}
\emph{Old}: Buttons above the main graph allowed to switch function so the main graph, where
only one could be active at the time. Some respondents expressed their disregard for the fact
that non-conflicting functions could not be active simultaneously. This warranted
concern results from the technical framework the \ac{FSS} is programmed in. 
\emph{New}: Buttons for switching functions are now located under the main graph, thereby
requiring less eye movement for accessing frequently used elements. They are grouped by their function from the user perspective (navigation versus editing)
instead of their function in the context of the \ac{FSS}'s backend programming. 

\paragraph{Bonus}
\emph{Old:} Participants learned that they would receive a bonus depending on the accuracy of their final forecasts.
Some respondents wanted to know how accuracy would be measured, and some asked to split the bonus across products. 
\emph{New:} Following the requests, now a bonus can be earned for each product depending on the relative
improvement over the statistical forecast of 0.20 \acs{USD} per percentage point improvement as measured
by the \ac{rMAE}, up to 1.00 \acs{USD} per product.

\paragraph{Framing of the Trend Component}
\emph{Old:} The trend of the Prophet model was depicted in the sales data and labeled as such. 
Many respondents were confused about its meaning and purpose because its slope was usually
flat but ``Trend'' is connotated with a clear upwards or downwards movement. 
One respondent accurately noted that the trend component, other than weekday, yearly, or event effects
is not tied to a tangible underlying reason but time itself and therefore confusing. 
\emph{New:} The trend line is only displayed when it is relevant and is introduced as 
``general level of sales''. 

\paragraph{Task Complexity}
\emph{Old:} To examine how the \ac{FSS} designs
interact with somewhat realistic perturbations, I would randomly select two product time series and introduce one of the following variations respectively:
First, one product time series was altered so that the last 16 weeks would have 30\% more sales on Mondays
and 20\% fewer sales on Tuesdays. The Prophet model and forecast were not updated so that the weekly component was severely misspecified for the forecast period. 
Participants were informed about recent market fluctuations on Mondays and Tuesdays, which they would have to identify and correct for. 
Second, the other time series would be subject to unfounded bad advice from an imaginary colleague. For
one forecast, he would advise altering the forecast based on his ``gut feeling''. That gut feeling in actuality reflects the additive Prophet model forecast, only that the weekly effect is multiplied by $-2$. 
These two variations were intended to observe how participants would interact with the three \ac{FSS}
designs under these variations. Further, the original intended forecasting horizon was 28 days to mimic the M5 competition. 
Some respondents in the Transparent treatment commented that they would find it very cumbersome to adjust each value manually (which they would not have to, of course).
\emph{New:} Because some pre-test participants reported being overwhelmed with the task complexity, I removed the above variations from the final experimental procedure and limited the forecasting horizon to 14 days. 

\paragraph{Visual Clutter and Performance}
\emph{Old:} The old design featured different colors for each element in the graph.
Historical values were visualized as scatter points, and a table depicted the forecast values
numerically. Some participants reported the \ac{FSS} to be ``overwhelming'' and had performance issues.
\emph{New:} 
The observed values are now shown in line format, mainly because this is less demanding on the computer running the \ac{FSS}. 
The graph depicts the model fit in gray and the observed values and forecast values in purple.
This creates a visual continuity of observed values and forecast, as          
the latter would optimally become the former while preserving the separateness of the forecast
by marking it with triangles. I completely removed the table because participants did not utilize it and its two-way synchronization with the graph turned out unreasonably resource intense.

\subsection{Results}

Due to \ac{mTurk}'s mechanics, each experimental treatment initially contains more than 80 participants.
I drop seven duplicate participants from \ac{T} who managed to participate in it after having participated in \ac{O}.
I further drop 14 participants with overall completion times below three minutes. Table \ref{tbl:participants}
provides the exact numbers.

\begin{table}[h]
\centering
\begin{tabular}{@{}cccc@{}}
\toprule
\multirow{2}{*}{\textit{Filter}}                                            & \multicolumn{3}{c}{\textbf{Treatment}} \\
                & \textbf{O} & \textbf{T} & \textbf{TA} \\ \midrule
Initial Sample  & 81         & 80         & 89          \\ \midrule
Drop Duplicates & 81         & 80         & 82          \\ \midrule
\begin{tabular}[c]{@{}c@{}}Drop Completion Time \\ below 3 min\end{tabular} & 75          & 77          & 78         \\ \bottomrule
\end{tabular}
\caption{Number of Participants per FSS treatment after respective filtering.}
\label{tbl:participants}
\end{table}

As described for Pre-Study 1, I resample the data to make comparisons between respective \ac{AV}
and \ac{rMAE} metrics less prone to be biased by differing distributions of product time series among treatments.
The Figures \ref{fig:study3_appendix_resampling} in the Appendix \ref{appendix} visualize the trade-off. Again, making
the count of time series even between treatments is worth the over-and underrepresentation of some
participants. It makes \ac{rMAE} and \ac{AV} metrics more representative to allow
for comparisons. The results do not change qualitatively through resampling.

\subsubsection{Hypotheses Testing}
\begin{figure}[h]
	\includegraphics[width=1.0\textwidth]{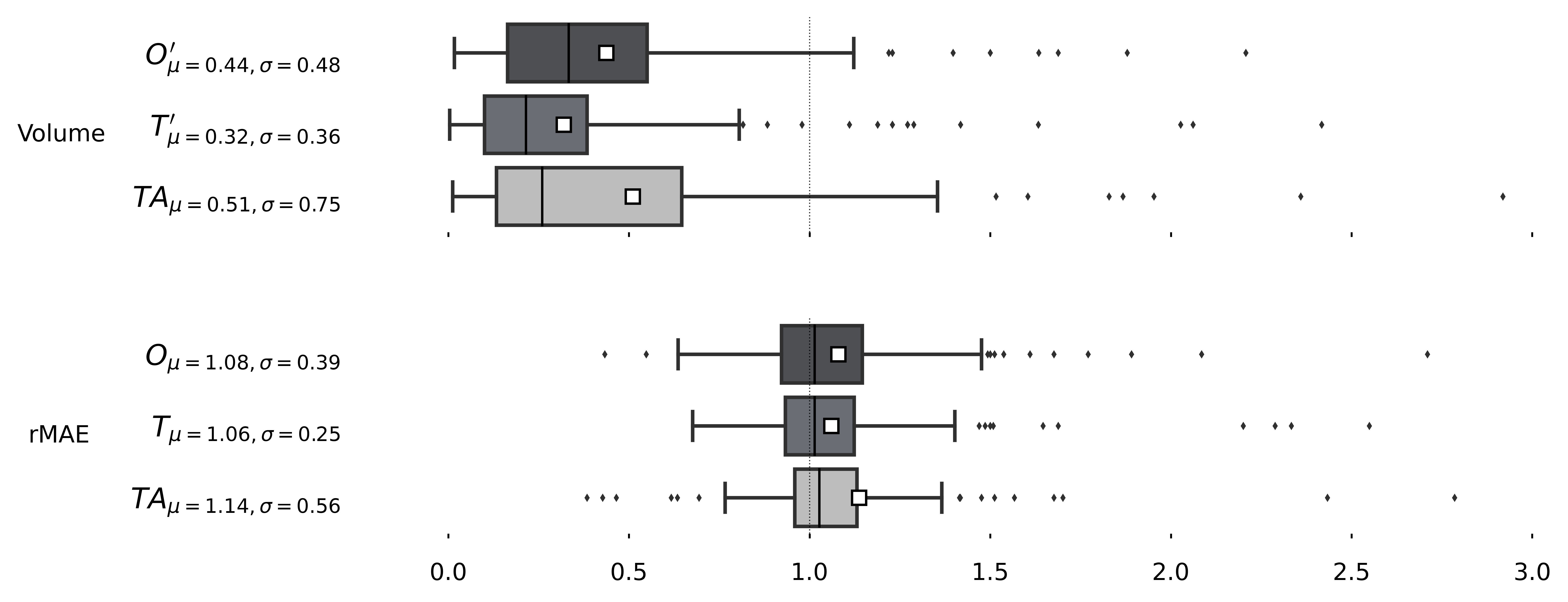}
	\caption{$AV$ and $rMAE$ per FSS design conditional on positive $AV$. White squares mark the mean values.}
	\label{fig:study3_volumermae_nonzero}
\end{figure}

\paragraph{Hypothesis 1a: Adjustment Volume}
The average \ac{AV} per treatment is shown in Table \ref{tbl:study3_unconditional}.
Both transparent conditions (\ac{T} and \ac{TA}) have a significantly lower \ac{AV}
than the Opaque (\ac{O}). This finding may be driven by the lower \ac{AF} (see Equation \ref{eq:adjfreq}) in \ac{T} and \ac{TA}. Participants
in \ac{O} leave only 13.6\% of forecasts untouched whereas those in \ac{TA} do not adjust 38\% of forecasts at all.
Drilling down on the \ac{AV} conditional on non-zero adjustments, Figure \ref{fig:study3_volumermae_nonzero}
depicts the distributions of \ac{AV}s.
As expected, the conditional \ac{AV}s are larger than the unconditional \ac{AV}s.
Now, only the \ac{AV} of \ac{T} is significantly smaller than that of \ac{O}. 
Strikingly, \ac{TA}'s conditional \ac{AV} has the largest mean and standard deviation. 
These results confirm Hypothesis \ref{hyp:1a} with the strong caveat that \emph{if} participants adjust forecasts, their \ac{AV}
in \ac{TA} is larger and highly variant. 

\begin{table}[h]
\centering
\begin{tabular}{llll}
\toprule
{} & \multicolumn{2}{c}{Adj. Volume} & Adj. Freq. \\
{} &  Mean &   Std & \\
\ac{FSS} &       &       & \\
\midrule
\ac{O}    &  0.38 &  0.47 &0.86\\
T    &  0.25 &  0.34 &0.78\\
TA   &  0.32 &  0.64 &0.62\\
\bottomrule
\end{tabular}
\caption{Unconditional $AV$ and $AF$ per treatment}
\label{tbl:study3_unconditional}
\end{table}

\paragraph{Hypothesis 1b: Forecast Quality}

Table \ref{tbl:study3_unconditional_rmae} provides the unconditional \ac{rMAE} with $rMAE = \frac{MAE^{final}}{MAE^{model}}$ averaged per \ac{FSS} design. There is no signifcant difference considering the mean values. As with \ac{AV}, however, \ac{rMAE} is most variable in \ac{TA}
and least variable in \ac{T}. 

\begin{table}[h]
\centering
\begin{tabular}{lll}
\toprule
{} &  Mean &   Std \\
Mode &       &       \\
\midrule
\ac{O}    &  1.07 &  0.36 \\
\ac{T}    &  1.05 &  0.22 \\
\ac{TA}   &  1.08 &  0.44 \\
\bottomrule
\end{tabular}
\caption{Unconditional $rMAE$ per treatment}
\label{tbl:study3_unconditional_rmae}
\end{table}

Examining the \ac{rMAE} scores conditional on positive \ac{AV}, Figure \ref{fig:study3_volumermae_nonzero}
makes clear that participants in \ac{TA} have the greatest variance in their forecasting accuracy with many outliers \emph{if} they adjust forecasts.
A one-way ANOVA, however, finds no significant differences in group means (p=0.54) after logarithmizing
the initially log-normally distributed \ac{rMAE} values. 
Thereby, I reject Hypothesis \ref{hyp:1b} but point out that the Transparent design shows the lowest error rates and standard deviations whereas the Transparently Adjustable design has the highest respective values. 

\paragraph{Hypothesis 2: User Satisfaction}

I measure user satisfaction using the first four Likert items at the end of the experiment, which ask participants to self-assess their understanding of the sales model, its usefulness, their ability to bring in intuition, and their satisfaction with their final forecasts. The fifth item inquires whether the bonus was the primary motivator for adjustments. Aggregate results are summarised in Appendix \ref{appendix} Table \ref{tab:study3_appendix_answers}.

Participants in the \ac{TA} condition report a significantly lower understanding of the model than those in the other conditions (p = 0.004). The ability to incorporate intuition is rated significantly higher in \ac{O} compared to \ac{TA} (p = 0.002), and participants in \ac{O} are more satisfied with their final forecasts than those in \ac{T} and \ac{TA} (p < 0.001). Additionally, participants in \ac{O} indicate their adjustments are significantly more motivated by the bonus than participants in \ac{TA}. While other differences are not statistically significant, they suggest a general dissatisfaction with \ac{TA} and satisfaction with \ac{O}.

I aggregate the first four items into a \emph{Satisfaction} scale. The Satisfaction scores for \ac{O}, \ac{T}, and \ac{TA} are 5.13, 4.74, and 4.31, respectively, with the difference between \ac{O} and \ac{TA} being statistically significant (p = 0.001). Therefore, Hypothesis~\ref{hyp:2} must be rejected given the data.

\subsubsection{Data Exploration}

\begin{figure}[h]
	\includegraphics[width=1.0\textwidth]{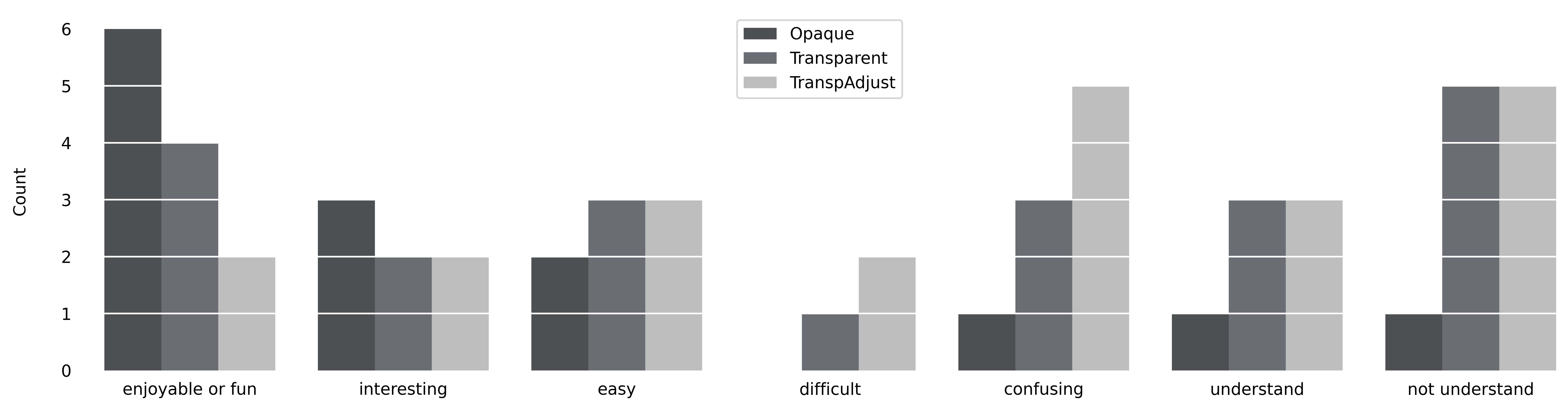}
	\caption{Counts of voluntary free-text comments.}
	\label{fig:study3_comments}
\end{figure}

\paragraph{Comments}
At the end of the \ac{HIT}, a textbox asks participants whether they particularly liked or disliked anything.
About half of the participants in each condition left a comment.
Comments in \ac{O} are pretty uniform, thanking for the unique study, stating that it was fun or interesting,
and easy to control. Participants in \ac{T} and \ac{TA} give more diverse comments. The number of comments stating enjoyment
declines from \ac{O} to \ac{T} to \ac{TA}, while more comments suggest difficulties and confusion. 
Comments about understanding \emph{and} not understanding both increase in the transparent conditions.
A large number of comments in \ac{T} and \ac{TA} states that they see the value in the forecasting tool but wished
for more time, experience, or explanations to get used to it. 
Figure \ref{fig:study3_comments} provides the number of comments for a wordstem or meaning (i.e. ``confusing'' counts comments that include ``confused'').

Some participants indicate a professional background that relates to the \ac{HIT}. 
Three participants, two of which having worked in gastronomy, state that they would or would have liked to see
such a tool at work. 
Another participant who states having worked as a sales manager for beverages states that he found the \ac{HIT} ``amusing'' because
he used to run a similar model in his head when making orders.

\begin{figure}[h]
	\includegraphics[width=1.0\textwidth]{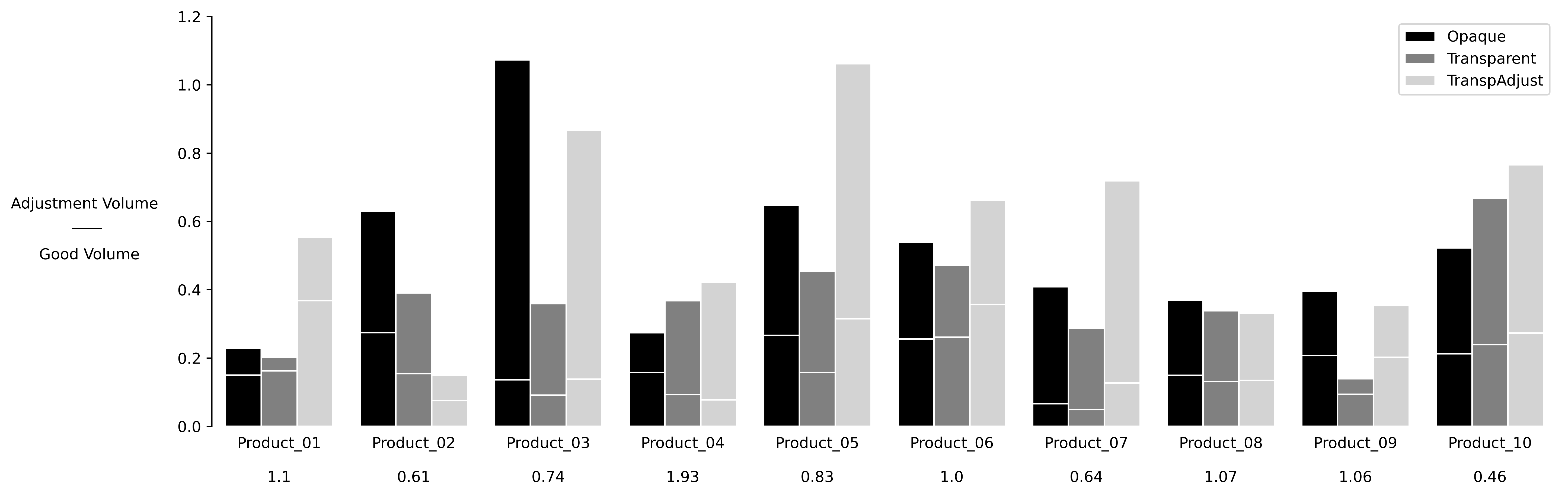}
	\caption{$Adjustment Volume$ and proportion of $good$ adjustments, marked by each bar's lower part, per product and FSS design. Values under the product index indicate Prophet's $rMAE$ relative to a simple exponential smoothing model.}
	\label{fig:study3_perproduct_volume}
\end{figure}

\begin{figure}[h]
	\includegraphics[width=1.0\textwidth]{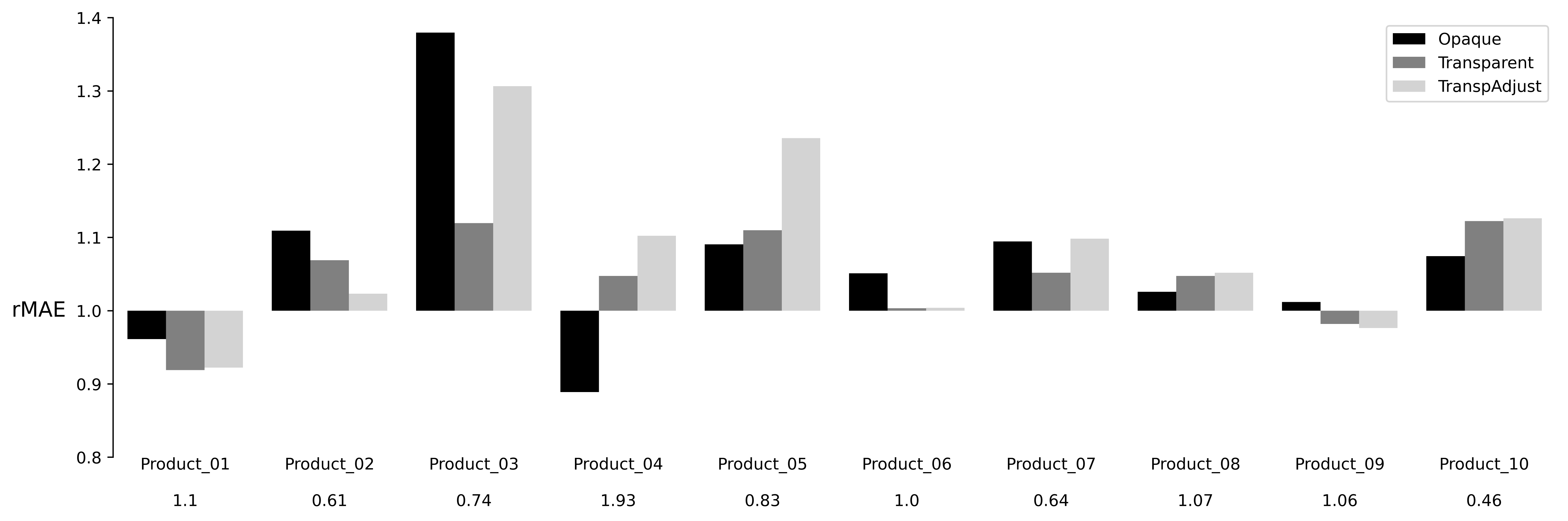}
	\caption{$rMAE$ per product time series and FSS design. Values under the product index indicate Prophet's $rMAE$ relative to a simple exponential smoothing model.}
	\label{fig:study3_perproduct_rmae}
\end{figure}

\paragraph{Individual Product Time Series}
Having established the general behavioral effects of transparency, I will now explore
how they relate to individual product time seris. Figure \ref{fig:study3_perproduct_volume}
shows the unconditional \ac{AV} per \ac{FSS} design and product time series and Figure \ref{fig:study3_perproduct_rmae}
gives the respective \ac{rMAE}s. 
The values below the product indices in both figures give Prophet's \ac{rMAE} relative to a \ac{SES} model.
Notably, average improvements through adjustments are conditional on Prophet performing worse than \ac{SES}.
Conversely, large disimprovements through adjustments only occur where Prophet is significantly more accurate than \ac{SES}.

It stands out that Product\_01's forecast is significantly enhanced through judgmental interventions. It is also a product time series
for which \ac{SES} provides a more accurate forecast. Figure \ref{fig:study3_appendix_ts} in Appendix \ref{appendix} depicts all product time series
from the Main Study, alongside the model forecast and the mean of adjusted final forecasts per treatment. Thereby, the
final forecast scatter points show a \emph{consensus adjustment}, that via the averaging of errors can be expected to be
more precise than any single forecast.
It is obvious that the model forecast for Product\_01 is subjectively, visually, \emph{and} actually too low.

\subsection{Discussion}
\label{discussion}

Did transparency achieve the expected positive effects on \ac{FSS} user interaction?
The experimental results offer much room for nuanced discussion and interesting new questions.
Transparency turns out to reduce overall adjustments and is highly appreciated by some
participants while others are overwhelmed and confused. The adjustment of individual components
turns out to be a highly risky strategy that worsens most forecasts.

\paragraph{Actual Satisfaction or Intimidation} In line with Hypothesis \ref{hyp:1a},
\ac{TA} participants refrain from any adjustments significantly more often than those in \ac{T} and \ac{O}.
A benevolent interpretation would be that their deeper engagement with the algorithm lets participants realize that judgmental interventions can not improve the model.
Put differently, granting more adjustment facilities conveys the algorithm's complexity which
intimidates users and prevents adjustments. 
This interpretation connects with the peculiar finding that \ac{TA} participants indicate a significantly lower
understanding of the model than \ac{TA} and \ac{O} participants. At first look, this result is puzzling since \ac{TA} receive the same 
information about the model as \ac{T} plus the adjustment options, which allow for intuitive
``learning by doing'' (as some comments indicate). Probably, the complexity of \ac{TA}
lets participants realize what they do not understand, hence leading to lower self-reported understanding.
This is supported by the highest self-reported understanding in \ac{O}, the treatment which does not
provide any explanations at all about the time series model. In general,
participants in \ac{O} display an ``ignorance is bliss'' attitude as they know the least but think they know the most and adjust accordingly.

There are two possibilities by which the low \acl{AF} in \ac{TA}, especially when
compared with \ac{O}, results from experimental artifacts.

\paragraph{Demand Effects} Firstly, a simple demand effect might be at play: Participants in \ac{T} are asked to comprehend
the statistical model, those in \ac{TA} additionally must comprehend the more complex user interface.
It might well be the case that participants in \ac{TA} perceive that comprehending the interface
and model makes up a significant portion of their \ac{HIT} so that they feel less inclined to adjust every time series forecast. Conversely, participants in \ac{O} might perceive
an obligation to adjust forecasts since this is the only tangible action they can take.
This interpretation implies that a significant proportion of participants is not sufficiently motivated
by the bonus.

\paragraph{Self-Selection} Secondly, the studies were launched simultaneously on mTurk using cloudresearch's survey group.
Thereby, Workers who have participated in one \ac{HIT} of that survey group can not take any other \ac{HIT} of that group.
This feature, prima facie innocent and useful, could have led to biased sampling:
Sophisticated Workers use internet forums and tools to find high-paying \ac{HIT}s.
The \ac{O} \ac{HIT}, while having the same description as all other conditions, had the highest
median hourly pay because it does not include as many explanations as the other conditions.
Thereby, more sophisticated Workers might have self-selected into the \ac{O} \ac{HIT} - 
a concern supported by the casual observation that the corresponding \ac{HIT} finished much faster than the others. Rationally, experienced and sophisticated Workers easily identify the asymmetric incentive structure
and tend to perform more adjustments. Participants' significantly greater motivation through the bonus in O
compared to \ac{TA} is another cue for this kind of self-selection.

\paragraph{Variance and Skewness}
Once participants in \ac{TA} adjust a forecast, they tend to produce not only the largest average \acl{AV} and \ac{rMAE}.
but also those the highest standard deviations and skewness.
From the top five largest \ac{AV} per participant and product, four come from \ac{TA} and all
created their adjustments only by adjusting components, not single values (less than 1\%).
Obviously, redrawing a line or dragging the value for a weekly effect easily creates huge adjustments alongside high \ac{rMAE}s.
The skewness and high standard deviations of \ac{AV} and \ac{rMAE} in \ac{TA} alongside some suggestive comments point to the fact
that the variety and potency of tools in \ac{TA} was overwhelming for some participants. 
Put differently, \ac{TA} introduces \emph{more} features that introduce more
room for extreme reactions. These extreme reactions lead to highly variable and
unfavorable performance and satisfaction metrics in \ac{TA}.
Notably, the \emph{adjustable} part in \ac{TA}, not the \emph{transparent} part, seems to be at fault here:
The Transparent treatment (\ac{T}) has the lowest \ac{AV}s and \ac{rMAE} and the lowest standard deviations for these metrics.
The adjustment possibilities in \ac{TA} have trumped any positive effects of transparency observable in \ac{T}. 

\paragraph{The present study as a worst case scenario}
Admittedly, there is something schizophrenic about the experimental design of this study and similar previous studies.
If participants have no extra information, they, almost by definition, can not best a statistical model.
As one participant commented, the forecasts looked fine to her, and, in practice, she would sign them off as they are.
Maximum trust in the algorithm would therefore be characterized by no adjustments at all.
No engagement with the task would look the same. It is not trivial to differentiate the two,
as attention checks can easily be misspecified.
This experiment incentivized adjustments because some organizational environments may do the same as I have 
argued in Section \ref{incentivescheme}.
In a way however, the experimental design established a \emph{worst case} setting for \ac{FSS} interactions:
Without extra information, adjustments were set up to fail, however, they were encouraged by the bonus. 
Further, \ac{TA} introduced complex functionalities that explicitly overwhelmed some participants.
\emph{Best case} settings will almost certainly produce different results - their experimental operationalization, however, requires significantly more resources than the scope of this work could provide.
A \emph{best case} operationalization would involve experienced managers as participants, learning facilities with 
the \ac{FSS}, realistic extra information, and realistic incentive structures.

%% file: text/6-Conclusion.tex
\section{Conclusions}
\label{conclusions}

This study explored the impact of algorithmic transparency on forecast outcomes and user satisfaction. Although transparency did not result in significantly more accurate forecasts or higher user satisfaction, it led to fewer and less frequent adjustments, with the lowest variance in forecast errors. Conversely, the transparently adjustable \ac{FSS}, despite reducing adjustment frequency, resulted in more extreme adjustments and higher forecast error variance.

Self-reported understanding and participant feedback revealed a mixed response: while some appreciated the transparency, many felt overwhelmed by the additional complexity, particularly when paired with component-wise adjustment options. This highlights the need for comprehensive training in transparent and interactive systems to avoid information overload and ensure meaningful interaction between the user and the system.

In summary, algorithmic transparency can foster user appreciation and stable adjustments but also poses challenges for untrained users, leading to variability in forecasts. Careful management of powerful adjustment tools is essential, as they can lead to inaccurate outcomes without proper guidance. Therefore, the introduction of transparent \ac{FSS} should be accompanied by extensive training to maximize its benefits, ultimately making algorithmic transparency a valuable tool for enhancing user understanding and more informed decision-making in time series forecasting.

%% file: text/x-appendix.tex
\section{Appendix}
\label{appendix}

\begin{table}[ht]
\centering
\caption{Agreement on Statements per FSS design. The * symbol indicates a significant difference from both other treatments.}
\label{tab:study3_appendix_answers}
\begin{tabular}{lccc}
\hline
\textbf{Category} & \textbf{Treatment} & \textbf{Average ($\mu$)} & \textbf{Std. Dev. ($\sigma$)} \\
\hline
Understanding      & O         & 4.44  & 1.74 \\
                   & T         & 4.57  & 1.82 \\
                   & TA        & 3.70* & 1.69 \\
\hline
Usefulness         & O         & 5.64  & 1.34 \\
                   & T         & 5.35  & 1.53 \\
                   & TA        & 5.13  & 1.52 \\
\hline
Bring in Intuition & O         & 5.21* & 1.44 \\
                   & T         & 4.57  & 1.90 \\
                   & TA        & 4.23  & 1.74 \\
\hline
Satisfaction       & O         & 5.24* & 1.39 \\
                   & T         & 4.46  & 1.93 \\
                   & TA        & 4.16  & 1.97 \\
\hline
Bonus Motivation   & O         & 4.81* & 1.85 \\
                   & T         & 4.14  & 1.89 \\
                   & TA        & 3.81  & 2.08 \\
\hline
\end{tabular}
\end{table}

\begin{figure}[h]
 	\begin{subfigure}[t]{0.5\textwidth}
 		\centering
		\includegraphics[width=1.0\textwidth]{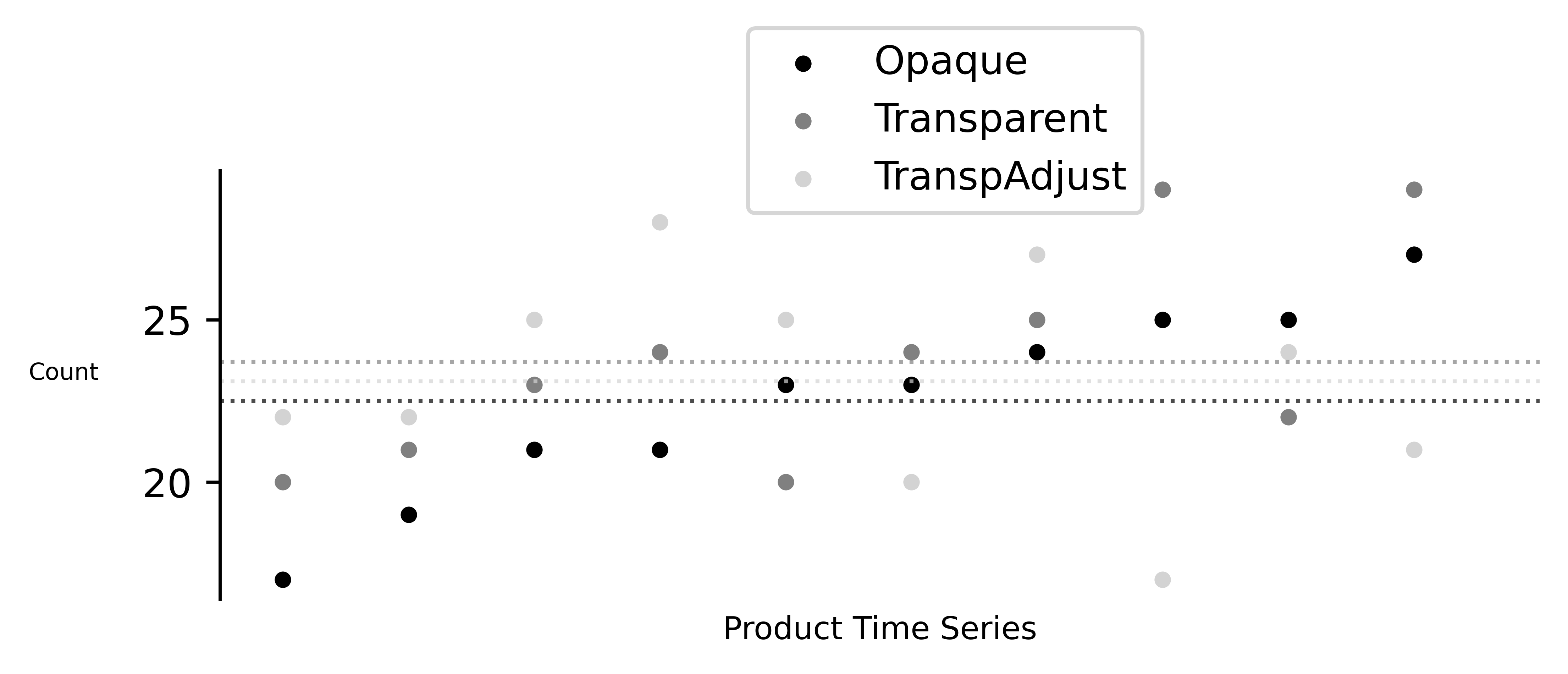}
		\caption{Imbalanced product time series representation in the original data. Dashed lines indicate the optimal counts.}
		\label{fig:study3_appendix_resampling_prod}
 	\end{subfigure}%
 	\begin{subfigure}[t]{0.50\textwidth}
		\includegraphics[width=1.0\textwidth]{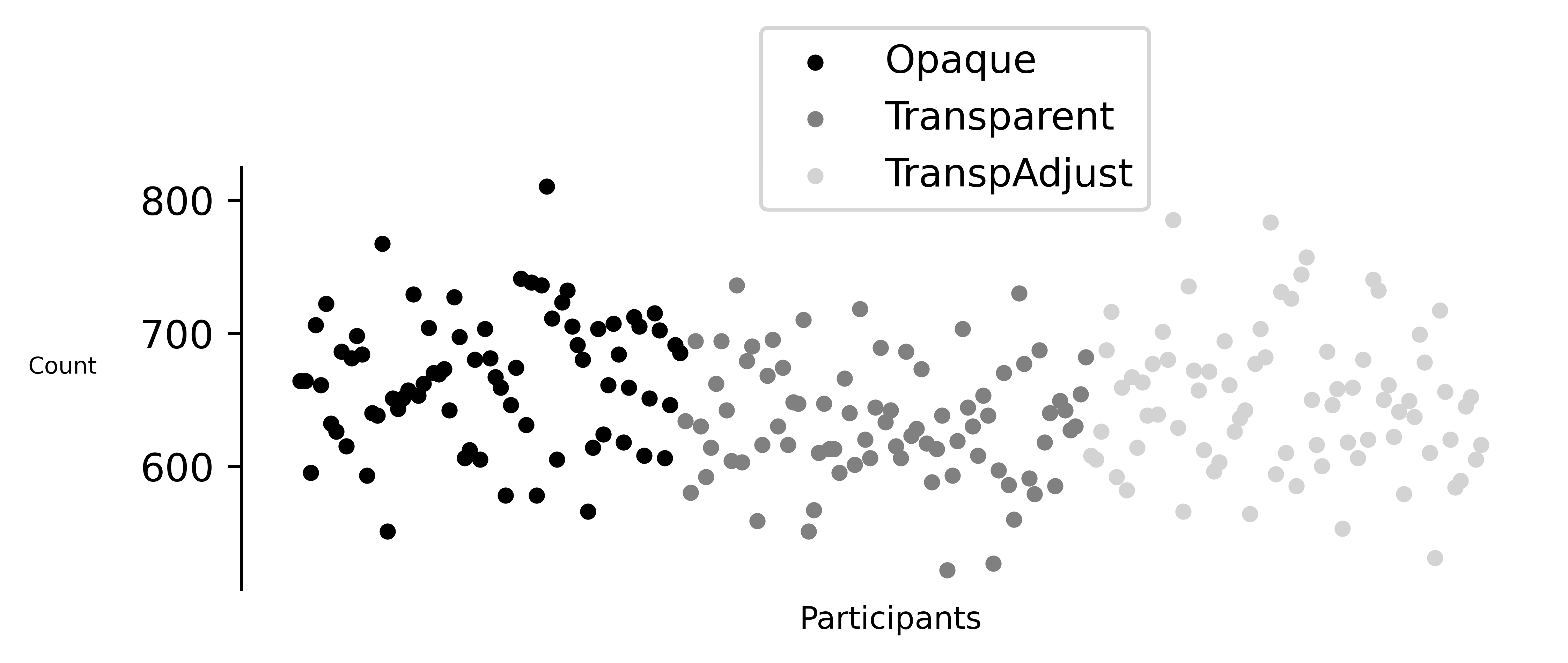}
		\caption{Imbalanced participant representation in the resampled data.}
		\label{fig:study3_appendix_resampling_part}
 	\end{subfigure}
 	\caption{Effects of resampling on product and participant counts.}
 	\label{fig:study3_appendix_resampling}
 \end{figure}
 \clearpage

	\begin{figure}[h]
		\includegraphics[width=1.0\textwidth]{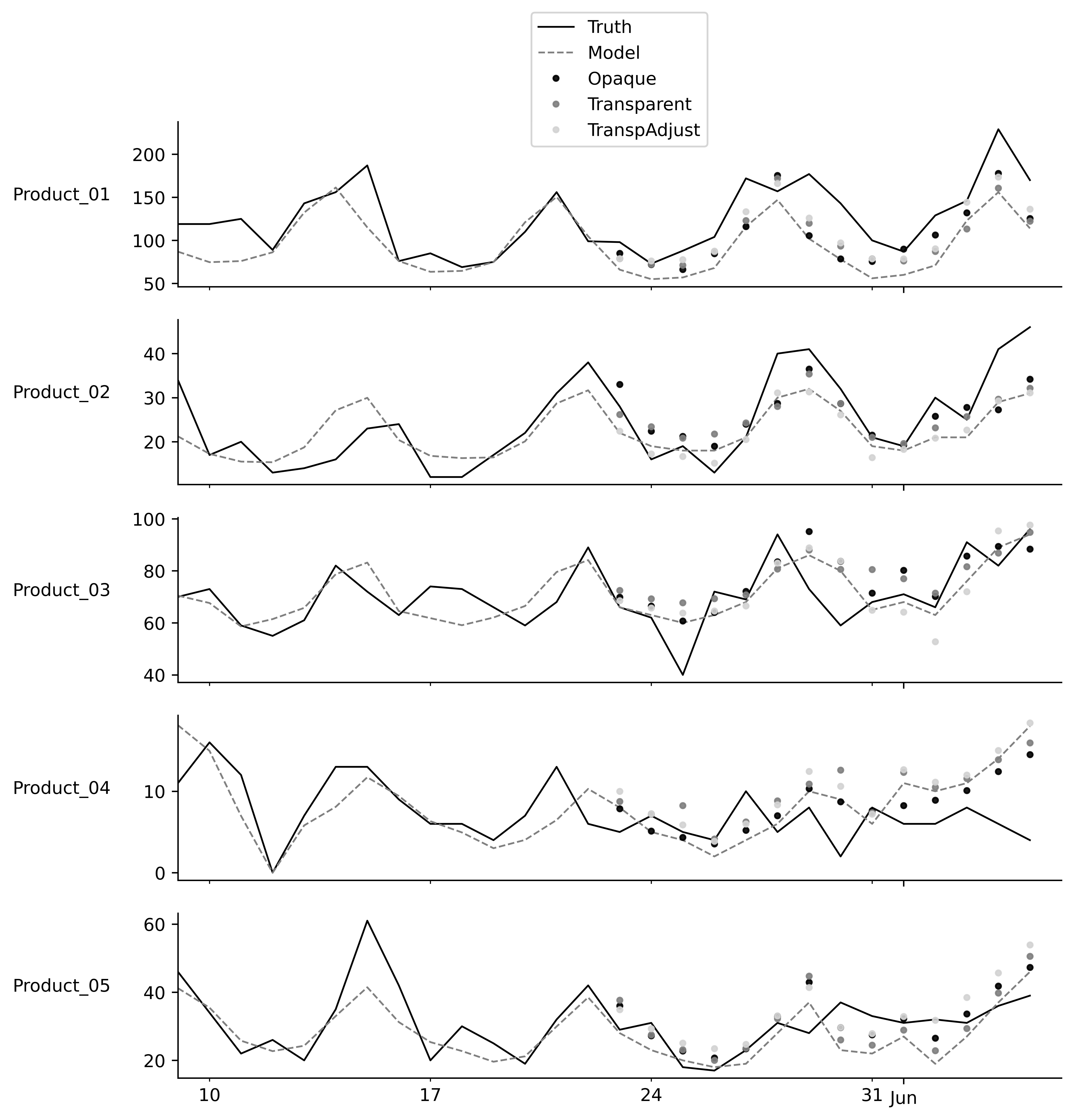}

		\label{fig:stud3_app1}
	\end{figure}
	\clearpage 
 	\begin{figure}[h]
		\includegraphics[width=1.0\textwidth]{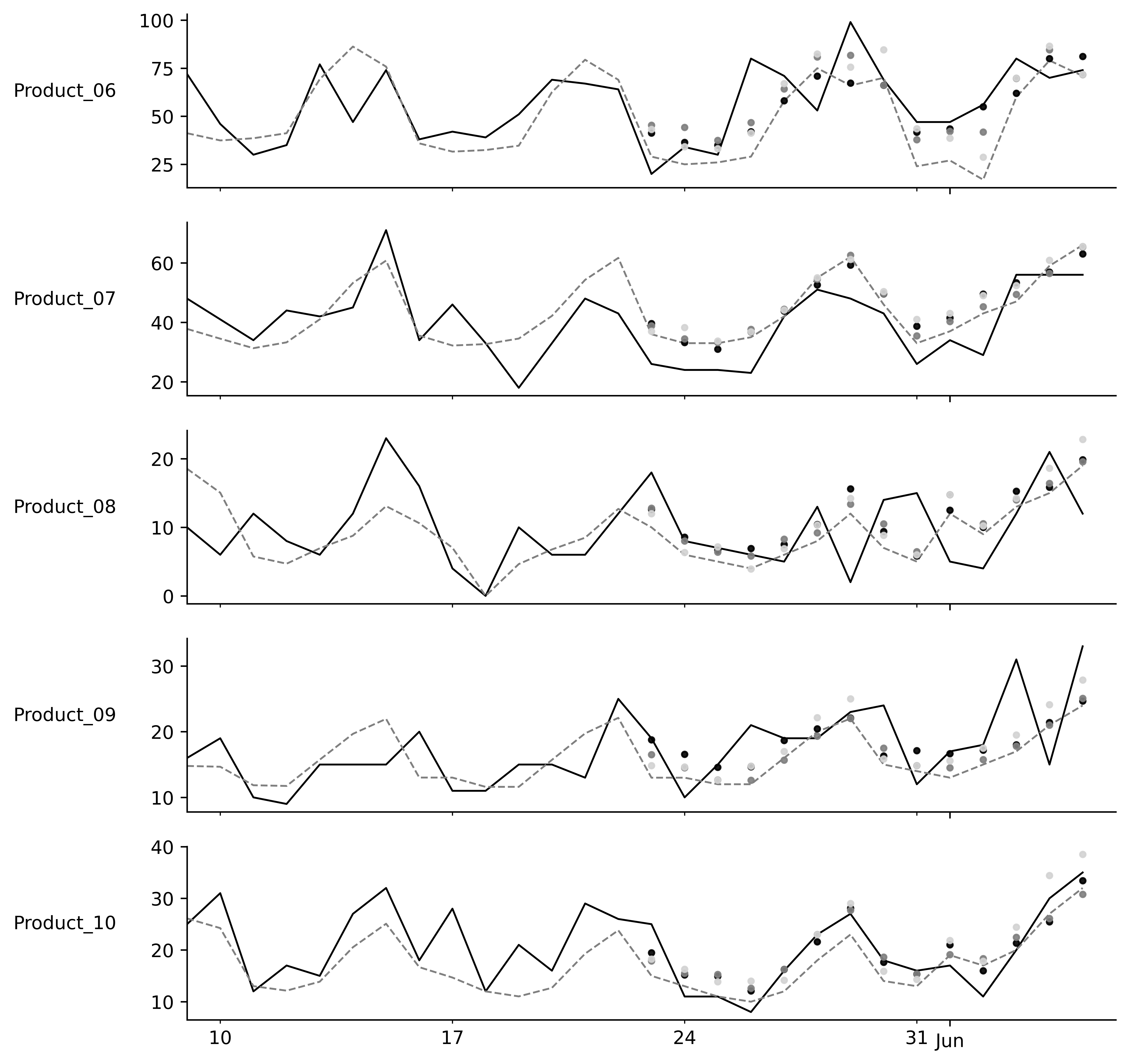}
		\caption{Product time series with model forecasts and respective means of final forecasts per FSS design conditional on positive adjustments.}
		\label{fig:study3_appendix_ts}
 	\end{figure}